\documentclass[10pt,letterpaper]{article}
\usepackage{opex3}
\usepackage{amsmath}
\usepackage{amssymb}

\begin{document}
\title{Calculating the Fine Structure of a Fabry-Perot Resonator using Spheroidal Wave Functions}
\author{M. Zeppenfeld$^1$ and P.W.H. Pinkse$^{1,2}$}
\address{$^1$Max-Planck-Institut f\"ur Quantenoptik,  Hans-Kopfermann-Str. 1, D-85748 Garching, Germany\\
$^2$Mesa+ Institute for Nanotechnology, University of Twente, PO Box 217, 7500 AE Enschede, The Netherlands}
\email{email: martin.zeppenfeld@mpq.mpg.de}


\begin{abstract*}
A new set of vector solutions to Maxwell's equations based on solutions to the wave equation in spheroidal coordinates allows laser beams to be described beyond the paraxial approximation. Using these solutions allows us to calculate the complete first-order corrections in the short-wavelength limit to eigenmodes and eigenfrequencies in a Fabry-Perot resonator with perfectly conducting mirrors. Experimentally relevant effects are predicted. Modes which are degenerate according to the paraxial approximation are split according to their total angular momentum. This includes a splitting due to coupling between orbital angular momentum and spin angular momentum.\\
\end{abstract*}


\bibliographystyle{unsrt}

\section{Introduction}
From cavity quantum electrodynamics to deterministic single-photon generation, a host of exciting recent experiments are made possible due to the precise control of light achievable inside optical resonators~\cite{Schuster08,Kuhn02}. In the case of Fabry-Perot resonators, the detailed understanding of the resonator modes necessary for these experiments is generally based on solutions to the paraxial wave equation. Despite being approximate in nature, these solutions allow many crucial features of the resonator modes to be explained, from the Gaussian profile of the lowest-order TEM$_{0,0}$ mode to the regularly spaced eigenmode spectrum.

Despite its success, novel experiments allow the boundary of the validity of the paraxial approximation to be probed. This includes microresonators, where the small mode volume leads to a violation of the assumed paraxiality, and ultra-high-finesse resonators, where the high spectral resolution allows even minute details to be resolved. Understanding all observable features in such experiments requires the electromagnetic field to be treated more precisely than the paraxial approximation allows.

In the past, several approaches have been used to obtain corrections to the paraxial approximation for resonators~\cite{Lazutkin68,Laabs99,Visser05,Zomer07}. Lazutkin~\cite{Lazutkin68} chose an analytic expansion of the wavefunction to satisfy the wave equation with appropriate boundary conditions. Laabs \emph{et al.}~\cite{Laabs99} calculated a resonator round-trip propagation matrix using the method of Lax \emph{et al.}~\cite{Lax75} of calculating corrections to the paraxial equation. The effect of spherical aberrations was calculated by Visser \emph{et al.}~\cite{Visser05} based on perturbation to Gaussian propagation inside a resonator. Zomer \emph{et al.}~\cite{Zomer07} used a diffraction integral to propagate light around a resonator.

None of these papers take into account all effects needed to calculate the complete first-order correction beyond the paraxial approximation of the electromagnetic eigenfrequencies of a resonator. Lazutkin and Zomer \emph{et al.} restricted their analysis to scalar fields in two-dimensional resonators. Visser \emph{et al.} focused solely on spherical aberrations, albeit leading to an analytical result reproduced here. The most rigorous approach so far by Laabs \emph{et al.} used an insufficient approximation by treating the mirror surfaces purely in terms of a position-dependent phase shift.

\begin{figure}[t]
\center
\includegraphics[width=0.5\textwidth]{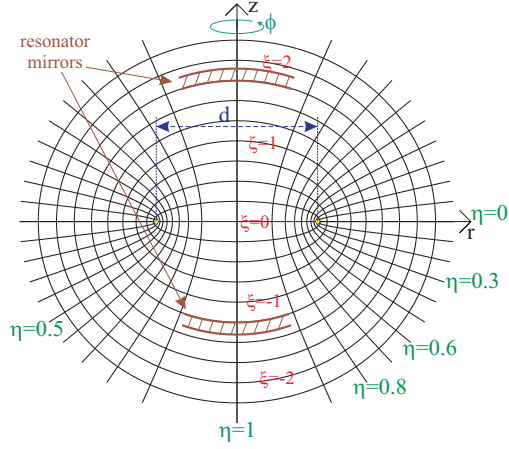}
\caption{The oblate spheroidal coordinate system. The $\xi$-coordinate describes the set of ellipses with a common pair of focal points separated by a distance $d$, the $\eta$-coordinate describes the set of hyperbolas with the same two focal points, and the $\phi$-coordinate describes the rotational angle around the z-axis. A pair of mirrors forming a resonators can be matched to the coordinate system as indicated, in this case with $\xi_-=-1.25$ and $\xi_+=1.75$}\label{Coordinates}
\end{figure}

An alternative approach for treatment of the resonator eigenmode problem is through the use of spheroidal coordinates. The spheroidal coordinate system, depicted in Fig.~\ref{Coordinates}, is ideally suited for calculations with Gaussian beams due to the match between surfaces of constant phase of a beam and surfaces of constant $\xi$ as well as between the variation in beam diameter along a beam and surfaces of constant $\eta$. With appropriate approximations, these coordinates have been applied in the past to resonators to obtain results in agreement with paraxial theory \cite{Zimmerer63,Specht65}.

In this paper, spheroidal coordinates are used to calculate the eigenfrequencies of a Fabry-Perot resonator to first order beyond the paraxial approximation in the short-wavelength limit. The electric field inside the resonator is expanded in terms of vector spheroidal wave functions. Requiring the boundary conditions for perfectly conducting mirrors to be satisfied allows the expansion coefficients of the outgoing wave at each mirror to be expressed in terms of the expansion coefficients of the incoming wave. The resulting round-trip propagation matrix for the expansion coefficients is solved via perturbation theory, resulting in a compact expression for the resonator round-trip phase shift, Eqs.~(\ref{lambda_J,sigma+,nu}) and (\ref{lambda_J,sigma-,nu-1}). The paper concludes with a discussion on the experimental implications of this result.




\section{Mathematical Foundations}
This paper relies heavily on the theory of spheroidal wave functions as found elsewhere~\cite{Flammer57,Zeppenfeld09}. In particular, the theory of spheroidal wave functions in the short-wavelength limit has recently been significantly expanded in Ref.~\cite{Zeppenfeld09}. Here we briefly summarize the most important definitions and results from this paper.

The mapping from spheroidal coordinates $(\xi,\eta,\phi)$ to cylindrical coordinates $(r,z,\phi)$ is given by
\begin{equation}
r=\frac{d}{2}\sqrt{(1-\eta^2)(1+\xi^2)}\hspace{1cm}\mbox{and}\hspace{1cm}z=\frac{d}{2}\eta\xi,
\end{equation}
the $\phi$-coordinate being the same in both coordinate systems and $d$ being the interfocal distance as shown in Fig.~\ref{Coordinates}. The scalar wave equation is separable in spheroidal coordinates, allowing the solutions to be written as the product of three functions depending only on $\xi$, $\eta$ and $\phi$, respectively. This leads to so-called scalar spheroidal wave functions $\psi_{m\nu}=R_{m\nu}(\xi)S_{m\nu}(\eta)e^{im\phi}$ with
\begin{equation}\label{R_mnu}
R_{m\nu}(\xi)=e^{i\bar{c}\xi}\frac{(1-i\xi)^{\nu+m/2}}{(1+i\xi)^{\nu+m/2+1}}\,r_{m\nu}(\xi)
\end{equation}
and
\begin{equation}\label{S_mnu}
S_{m\nu}(\eta)=\bar{c}^{m/2}(1-\eta^2)^\frac{m}{2}e^{-\bar{c}(1-\eta)}s_{m\nu}(x),
\end{equation}
labeled by the indices $m$ and $\nu$. Note that $m$ is simply the integer orbital angular momentum due to the $e^{im\phi}$ $\phi$-dependence. The variable $x=2\bar{c}(1-\eta)$, not to be confused with the Cartesian coordinate, is introduced to simplify calculations. The parameter $\bar{c}=kd/2$, with wavevector $k$, quantifies the scaling of the coordinate system relative to the wavelength. For short wavelengths compared to $d$, $\bar{c}$ is a large number, and the functions $r_{m\nu}(\xi)$ and $s_{m\nu}(x)$ can be expanded as asymptotic series in $\frac{1}{\bar{c}}$. To lowest order in $\frac{1}{\bar{c}}$, $r_{m\nu}(\xi)=1$ and $s_{m\nu}(x)=L_\nu^{(m)}(x)$ where $L_\nu^{(m)}(x)$ is a Laguerre polynomial.

Based on the scalar spheroidal wave functions, vector spheroidal wave functions $\mathbf{E}_{J\sigma\nu}^\pm$ propagating  in the $\pm\xi$ direction are defined which satisfy the wave equation as well as $\nabla\cdot\mathbf{E}_{J\sigma\nu}^\pm=0$. The $\mathbf{E}_{J\sigma\nu}^\pm$ can therefore be regarded as the electric field of a solution to Maxwell's equations in free space. With spin angular momentum $\sigma=\sigma^\pm$ corresponding to left and right circular polarization, $J=m\pm1$ denotes the total integer angular momentum of the field about the symmetry axis.

\section{Satisfying the Boundary Conditions}
The electric field inside the resonator is expanded in terms of vector spheroidal wave functions as
\begin{equation}\label{linear combination of modes}
\mathbf{E}=\sum_{J\sigma\nu}(b_{J\sigma\nu}^+\mathbf{E}_{J\sigma\nu}^++b_{J\sigma\nu}^-\mathbf{E}_{J\sigma\nu}^-),
\end{equation}
with expansion coefficients $b_{J\sigma\nu}^\pm$. For perfectly conducting mirrors, the component of the electric field parallel to a mirror surface $\mathbf{S}$ must vanish at $\mathbf{S}$, i.e. $\mathbf{E}|_{\mathbf{S},||}=0$.

The mirror surfaces are matched to the spheroidal coordinate system as exemplified in Fig.~\ref{Coordinates} as follows. For a resonator of length $L$ with spherical mirrors with radius of curvature $R_-$ and $R_+$, the spheroidal coordinate system is appropriately scaled by choosing the distance $d$ between the focal points to be~\cite{Zimmerer63}
\begin{equation}\label{d=}
d=\sqrt{\frac{4L(R_++R_--L)(R_+-L)(R_--L)}{(R_++R_--2L)^2}}.
\end{equation}
In this case, the two surfaces of constant $\xi=\xi_\pm$ with
\begin{equation}\label{xi_pm=}
\xi_\pm=\pm\frac{2L(R_\mp-L)}{d\,(R_++R_--2L)}
\end{equation}
are separated by a distance $L$ and have a radius of curvature of $R_\pm$ on the resonator axis.

For a spherical mirror, a correction to the spheroidal surface of constant $\xi$ is generally necessary. Switching briefly to cylindrical coordinates $(z,r,\phi)$, the mirror surfaces are specified as the distance $z=\bar{z}_\pm(r,\phi)$ above the plane $z=0$. For a smooth, cylindrically symmetric mirror, $\bar{z}_\pm(r,\phi)$ can be expanded in powers of $r^2$ as
\begin{equation}\label{cylindrical mirror surface}
\bar{z}_\pm(r,\phi)=z_\pm\mp\frac{r^2}{2R_\pm}\mp c_{4\pm}r^4+{\cal O}(r^6).
\end{equation}
Here, $z_\pm$ and $c_{4\pm}$ are expansion coefficients and $R_\pm$ is the radius of curvature of the surface at $r=0$. For a spherical surface, $c_{4\pm}=1/(8R_\pm^3)$. Other values of $c_{4\pm}$ can be chosen to describe cylindrically-symmetric mirror aberrations. For example, for a spheroidal surface of constant $\xi=\xi_\pm$ we have $z_\pm=d\xi_\pm/2$ and $c_{4\pm}=1/(4d\,\xi_\pm R_\pm^2)$. For a parabolic mirror, $c_{4\pm}=0$ by definition.

Switching back to spheroidal coordinates, the mirror surfaces can equivalently be specified as the value of $\xi$ as a function of $x=2\bar{c}(1-\eta)$ and $\phi$. Eq.~(\ref{cylindrical mirror surface}) is transformed into
\begin{equation}\label{spheroidal mirror surface}
\xi\Big|_{\mathbf{S}_\pm}=\bar{\xi}_\pm(x,\phi)=\xi_\pm\mp f_{4\pm}\frac{x^2}{\bar{c}^2}+{\cal O}\left(\frac{x^3}{\bar{c}^3}\right)
\end{equation}
with
\begin{equation}
f_{4\pm}=\frac{d\,\xi_\pm^2R_\pm^2}{2}c_{4\pm}\mp\frac{\xi_\pm}{8},
\end{equation}
$\mathbf{S}_\pm$ being the mirror surface at $\xi_\pm$. Note that for small $r$, $x$ is proportional to $\bar{c}\,r^2$. Additionally, due to the factor $e^{-x/2}$ in the definition of $S_{m\nu}(\eta)$, Eq.~(\ref{S_mnu}), the spheroidal wave functions vanish for $x\gg1$, motivating the use of $x$ in Eq.~(\ref{spheroidal mirror surface}). The term linear in $x$ is missing due to the specific choice of $d$ and $\xi_\pm$ in Eqs. (\ref{d=}) and (\ref{xi_pm=}). As will be seen, the $x^2$ term in Eq.~(\ref{spheroidal mirror surface}) is the highest-order term in $x$ which must be retained to calculate first-order corrections to the resonator eigenfrequencies.

The mirror surfaces can now be parameterized in terms of the transverse coordinates $\eta$ and $\phi$ as
\begin{equation}
\mathbf{S}_\pm(\eta, \phi)=\mathbf{x}\Big(\bar{\xi}_\pm\big(x(\eta),\phi\big),\eta,\phi\Big)
\end{equation}
where $\mathbf{x}(\xi,\eta,\phi)$ is the vector in $\mathbb{R}^3$ from the origin to the point denoted in spheroidal coordinates by $(\xi,\eta,\phi)$. A basis for the tangent space to the mirror surface at a point on the mirror surface, needed to calculate the component of $\mathbf{E}$ parallel to the mirror surface, is given by
\begin{equation}\label{exact tangent space basis}
\left\{\frac{\partial\mathbf{S}_\pm}{\partial\phi}=h_\phi\mathbf{\hat{e}}_\phi, \frac{\partial\mathbf{S}_\pm}{\partial\eta}=h_\eta\mathbf{\hat{e}}_\eta+h_\xi\mathbf{\hat{e}}_\xi\frac{\partial\bar{\xi}_\pm}{\partial x}\frac{\partial x}{\partial\eta}\right\},
\end{equation}
where the $\mathbf{\hat{e}}_{u_i}$ are unit vectors and the $h_{u_i}$ are scale factors in spheroidal coordinates~\cite{Abramowitz+Stegun}. The component of the electric field parallel to the mirror surface is proportional to the inner product of the electric field with these basis vectors.

We now show that as long as we are only interested in corrections to the resonator eigenfrequencies to first order in $\frac{1}{\bar{c}}$, we can use $\{\mathbf{\hat{e}}_\phi, \mathbf{\hat{e}}_\eta\}$ as an approximation for the basis (\ref{exact tangent space basis}). For $x={\cal O}(1)$, we have the following order of magnitudes,
\begin{equation}\begin{split}
h_\xi&={\cal O}(1),\hspace{2.8cm}h_\eta={\cal O}(\bar{c}^{1/2}),\hspace{1.2cm}h_\phi={\cal O}(\bar{c}^{-1/2}),\\
\mathbf{\hat{e}}_\xi\cdot\mathbf{E}_{J\sigma\nu}^\pm&={\cal O}(\bar{c}^{-1/2}),\hspace{1cm}\mathbf{\hat{e}}_\eta\cdot\mathbf{E}_{J\sigma\nu}^\pm={\cal O}(1),\\
\frac{\partial\bar{\xi}_\pm}{\partial x}&={\cal O}(\bar{c}^{-2}),\hspace{2.3cm}\frac{\partial x}{\partial\eta}={\cal O}(\bar{c}),
\end{split}\end{equation}
as can be seen from the corresponding definitions in Ref.~\cite{Zeppenfeld09}. As a result, the contribution to $\frac{\partial\mathbf{S}_\pm}{\partial\eta}\cdot\mathbf{E}$ of the $\mathbf{\hat{e}}_\xi\cdot\mathbf{E}$ term is of order $\bar{c}^{-2}$ compared to the contribution of the $\mathbf{\hat{e}}_\eta\cdot\mathbf{E}$ term and can therefore be neglected, justifying our simplified basis.

To calculate $\mathbf{\hat{e}}_\eta\cdot\mathbf{E}_{J\sigma\nu}^\pm$ and $\mathbf{\hat{e}}_\phi\cdot\mathbf{E}_{J\sigma\nu}^\pm$ we additionally need the following expressions,
\begin{equation}\begin{split}
\mathbf{\hat{e}}_\phi\cdot(\mathbf{\hat{x}}\pm i\mathbf{\hat{y}})=\pm i\,e^{\pm i\phi}\hspace{4cm}
&\mathbf{\hat{e}}_\phi\cdot\mathbf{\hat{z}}=0\\
\mathbf{\hat{e}}_\eta\cdot(\mathbf{\hat{x}}\pm i\mathbf{\hat{y}})=-\eta\sqrt{\frac{1+\xi^2}{\eta^2+\xi^2}}e^{\pm i\phi}\hspace{2.3cm}
&\mathbf{\hat{e}}_\eta\cdot\mathbf{\hat{z}}=\xi\sqrt{\frac{1-\eta^2}{\eta^2+\xi^2}}\\
\mathbf{\hat{e}}_\xi\cdot(\mathbf{\hat{x}}\pm i\mathbf{\hat{y}})=\xi\sqrt{\frac{1-\eta^2}{\eta^2+\xi^2}}e^{\pm i\phi}\hspace{2.6cm}
&\mathbf{\hat{e}}_\xi\cdot\mathbf{\hat{z}}=\eta\sqrt{\frac{1+\xi^2}{\eta^2+\xi^2}}
\end{split}\end{equation}
For first-order corrections, the expressions with $\mathbf{\hat{e}}_\xi$ are not needed, but have been included for completeness. Using previous expressions and with a significant amount of algebra, one obtains
\begin{equation}\begin{split}\label{phi_dot_+}
&\mathbf{\hat{e}}_\phi\cdot\mathbf{E}_{J\sigma^+\nu}^+\Big|_{\mathbf{S}_\pm}=i\,e^{i\phi}\psi_{J-1,\nu}\Big|_{\mathbf{S}_\pm}=\\
&x^\frac{J-1}{2}\left(1-\frac{Jx}{8\bar{c}}\right)e^{-x/2}e^{iJ\phi}R_{J-1,\nu}(\xi_\pm)\Bigg\{-iL_{\nu-1}^{(J)}+iL_\nu^{(J)}+\frac{1}{\bar{c}}\bigg[\mp f_{4\pm}x^2L_{\nu-1}^{(J)}\pm f_{4\pm}x^2L_\nu^{(J)}-\\
&\frac{i(\nu+J-1)(2\nu-1)}{8}L_{\nu-2}^{(J)}+\frac{i(2\nu^2+2J\nu-5\nu-2J+1)}{8}L_{\nu-1}^{(J)}+\\
&\frac{i(2\nu^2+2J\nu+5\nu+3J+1)}{8}L_\nu^{(J)}-\frac{i(\nu+1)(2\nu+2J+1)}{8}L_{\nu+1}^{(J)}\bigg]+{\cal O}\left(\frac{1}{\bar{c}^2}\right)\Bigg\},
\end{split}\end{equation}
\begin{equation}\begin{split}\label{phi_dot_-}
&\mathbf{\hat{e}}_\phi\cdot\mathbf{E}_{J\sigma^-\nu}^+\Big|_{\mathbf{S}_\pm}=-i\,e^{-i\phi}\psi_{J+1,\nu}\Big|_{\mathbf{S}_\pm}=\\
&x^\frac{J-1}{2}\left(1-\frac{Jx}{8\bar{c}}\right)e^{-x/2}e^{iJ\phi}R_{J+1,\nu}(\xi_\pm)\Bigg\{-i(\nu+J+1)L_\nu^{(J)}+i(\nu+1)L_{\nu+1}^{(J)}+\\
&\frac{1}{\bar{c}}\bigg[\mp f_{4\pm}x^2(\nu+J+1)L_\nu^{(J)}\pm f_{4\pm}x^2(\nu+1)L_{\nu+1}^{(J)}-\frac{i(\nu+J)(\nu+J+1)(2\nu+1)}{8}L_{\nu-1}^{(J)}+\\
&\frac{i(\nu+J+1)(2\nu^2+3\nu+J+2)}{8}L_\nu^{(J)}+\frac{i(\nu+1)(2\nu^2+4J\nu+5\nu+2J^2+6J+4)}{8}L_{\nu+1}^{(J)}-\\
&\frac{i(\nu+1)(\nu+2)(2\nu+2J+3)}{8}L_{\nu+2}^{(J)}\bigg]+{\cal O}\left(\frac{1}{\bar{c}^2}\right)\Bigg\},
\end{split}\end{equation}
\begin{equation}\begin{split}\label{eta_dot_+}
&\mathbf{\hat{e}}_\eta\cdot\mathbf{E}_{J\sigma^+\nu}^+\Big|_{\mathbf{S}_\pm}=\\
&i\,\mathbf{\hat{e}}_\phi\cdot\mathbf{E}_{J\sigma^+\nu}^+\Big|_{\mathbf{S}_\pm}+x^\frac{J-1}{2}\left(1-\frac{Jx}{8\bar{c}}\right)e^{-x/2}e^{iJ\phi}\frac{1}{\bar{c}}\Bigg\{\frac{i\xi_\pm}{\sqrt{1+\xi_\pm^2}}\Big[R_{J,\nu-1}(\xi_\pm)\Big((\nu+J-1)L_{\nu-2}^{(J)}-\\
&(2\nu+J-1)L_{\nu-1}^{(J)}+\nu L_\nu^{(J)}\Big)+R_{J\nu}(\xi_\pm)\left((\nu+J)L_{\nu-1}^{(J)}-(2\nu+J+1)L_\nu^{(J)}+(\nu+1)L_{\nu+1}^{(J)}\right)\Big]+\\
&\frac{\xi_\pm^2}{2(1+\xi_\pm^2)}R_{J-1,\nu}(\xi_\pm)\left[(\nu+J-1)L_{\nu-2}^{(J)}-(3\nu+2J-1)L_{\nu-1}^{(J)}+(3\nu+J+1)L_\nu^{(J)}-(\nu+1)L_{\nu+1}^{(J)}\right]\Bigg\},
\end{split}\end{equation}
and
\begin{equation}\begin{split}\label{eta_dot_-}
&\mathbf{\hat{e}}_\eta\cdot\mathbf{E}_{J\sigma^-\nu}^+\Big|_{\mathbf{S}_\pm}=\frac{1}{i}\,\mathbf{\hat{e}}_\phi\cdot\mathbf{E}_{J\sigma^-\nu}^+\Big|_{\mathbf{S}_\pm}+x^\frac{J-1}{2}\left(1-\frac{Jx}{8\bar{c}}\right)e^{-x/2}e^{iJ\phi}\times\\
&\frac{1}{\bar{c}}\Bigg\{\frac{i\xi_\pm}{\sqrt{1+\xi_\pm^2}}\Big[R_{J\nu}(\xi_\pm)(\nu+J+1)\left(-(\nu+J)L_{\nu-1}^{(J)}+(2\nu+J+1)L_\nu^{(J)}-(\nu+1)L_{\nu+1}^{(J)}\right)+\\
&R_{J,\nu+1}(\xi_\pm)(\nu+1)\left(-(\nu+J+1)L_\nu^{(J)}+(2\nu+J+3)L_{\nu+1}^{(J)}-(\nu+2)L_{\nu+2}^{(J)}\right)\Big]+\\
&\frac{\xi_\pm^2}{2(1+\xi_\pm^2)}R_{J+1,\nu}(\xi_\pm)\Big[-(\nu+J)(\nu+J+1)L_{\nu-1}^{(J)}+(\nu+J+1)(3\nu+J+2)L_\nu^{(J)}-\\
&(\nu+1)(3\nu+2J+4)L_{\nu+1}^{(J)}+(\nu+1)(\nu+2)L_{\nu+2}^{(J)}\Big]\Bigg\}.
\end{split}\end{equation}
The $L_\nu^{(J)}$ are Laguerre polynomials $L_\nu^{(J)}(x)$ with the dependence on $x$ being implicit. Heavy use of recursion relations among the Laguerre polynomials has been made. Note that we have evaluated the spheroidal wave functions at the mirror surfaces $\mathbf{S}_\pm$. For $\mathbf{E}_{J\sigma\nu}^-$, the field running in the other direction, the results are the same except that $\xi_\pm$ is replaced by $-\xi_\pm$ and $f_{4\pm}$ is replaced by $-f_{4\pm}$.

Combining Eqs.~(\ref{phi_dot_+}-\ref{eta_dot_-}) with Eq.~(\ref{linear combination of modes}), the boundary condition for $\mathbf{E}$ for the mirror at $\xi_\pm$ can be reformulated as a matrix equation for the coefficients $b_{J\sigma\nu}^\mp$ of the outgoing wave in terms of the coefficients $b_{J\sigma\nu}^\pm$ of the incoming wave. We begin by defining the set of vector functions $\mathbf{v}_{J\sigma\nu}$ as
\begin{equation}
\mathbf{v}_{J\sigma^+\nu}=x^\frac{J-1}{2}\left(1-\frac{Jx}{8\bar{c}}\right)e^{-x/2}\left(-L_{\nu-1}^{(J)}+L_\nu^{(J)}\right)(i\,\mathbf{\hat{e}}_\phi-\mathbf{\hat{e}}_\eta)e^{iJ\phi}
\end{equation}
and
\begin{equation}
\mathbf{v}_{J\sigma^-\nu}=x^\frac{J-1}{2}\left(1-\frac{Jx}{8\bar{c}}\right)e^{-x/2}\left((\nu+J+1)L_\nu^{(J)}-(\nu+1)L_{\nu+1}^{(J)}\right)(-i\,\mathbf{\hat{e}}_\phi-\mathbf{\hat{e}}_\eta)e^{iJ\phi},
\end{equation}
and by defining the functions $\mathbf{u}_{1J\sigma\nu}$ and $\mathbf{u}_{2J\sigma\nu}$ according to
\begin{equation}\begin{split}\label{u1+}
&\mathbf{u}_{1J\sigma^+\nu}=\sqrt{x}(\psi_{J,\nu-1}+\psi_{J\nu})\mathbf{\hat{e}}_\eta=\\
&\frac{1}{2}\,R_{J,\nu-1}(\xi)\bigg(\mathbf{v}_{J\sigma^-,\nu-2}-\mathbf{v}_{J\sigma^-,\nu-1}-(\nu+J-1)\mathbf{v}_{J\sigma^+,\nu-1}+\nu\mathbf{v}_{J\sigma^+\nu}\bigg)+\\
&\frac{1}{2}\,R_{J\nu}(\xi)\bigg(\mathbf{v}_{J\sigma^-,\nu-1}-\mathbf{v}_{J\sigma^-\nu}-(\nu+J)\mathbf{v}_{J\sigma^+\nu}+(\nu+1)\mathbf{v}_{J\sigma^+,\nu+1}\bigg),
\end{split}\end{equation}
\begin{equation}\begin{split}\label{u1-}
&\mathbf{u}_{1J\sigma^-\nu}=\sqrt{x}\Big((\nu+J+1)\psi_{J\nu}+(\nu+1)\psi_{J,\nu+1}\Big)\mathbf{\hat{e}}_\eta=\\
&\frac{1}{2}\,R_{J\nu}(\xi)(\nu+J+1)\bigg(\mathbf{v}_{J\sigma^-,\nu-1}-\mathbf{v}_{J\sigma^-\nu}-(\nu+J)\mathbf{v}_{J\sigma^+\nu}+(\nu+1)\mathbf{v}_{J\sigma^+,\nu+1}\bigg)+\\
&\frac{1}{2}\,R_{J,\nu+1}(\xi)(\nu+1)\bigg(\mathbf{v}_{J\sigma^-\nu}-\mathbf{v}_{J\sigma^-,\nu+1}-(\nu+J+1)\mathbf{v}_{J\sigma^+,\nu+1}+(\nu+2)\mathbf{v}_{J\sigma^+,\nu+2}\bigg),
\end{split}\end{equation}
\begin{equation}\begin{split}\label{u2+}
&\mathbf{u}_{2J\sigma^+\nu}=x\psi_{J-1,\nu}e^{i\phi}\mathbf{\hat{e}}_\eta=\\
&-\frac{x}{2}R_{J-1,\nu}(\xi)\mathbf{v}_{J\sigma^+\nu}-\frac{1}{2}R_{J-1,\nu}(\xi)(\mathbf{v}_{J\sigma^-,\nu-2}-2\mathbf{v}_{J\sigma^-,\nu-1}+\mathbf{v}_{J\sigma^-\nu})+{\cal O}\left(\frac{1}{\bar{c}}\right)
\end{split}\end{equation}
and
\begin{equation}\begin{split}\label{u2-}
&\mathbf{u}_{2J\sigma^-\nu}=x\psi_{J+1,\nu}e^{-i\phi}\mathbf{\hat{e}}_\eta=\\
&-\frac{x}{2}R_{J+1,\nu}(\xi)\mathbf{v}_{J\sigma^-\nu}-\frac{1}{2}R_{J+1,\nu}(\xi)[(\nu+J)(\nu+J+1)\mathbf{v}_{J\sigma^+\nu}-\\
&2(\nu+1)(\nu+J+1)\mathbf{v}_{J\sigma^+,\nu+1}+(\nu+1)(\nu+2)\mathbf{v}_{J\sigma^+,\nu+2}]+{\cal O}\left(\frac{1}{\bar{c}}\right).
\end{split}\end{equation}
The parallel component of the electric field of $\mathbf{E}_{J\sigma\nu}^+$ at the mirror surfaces can then be written in terms of the $\mathbf{v}_{J\sigma\nu}$ as
\begin{equation}\begin{split}\label{sigma+ on mirror}
&\mathbf{E}_{J\sigma^+\nu}^+\Big|_{\mathbf{S}_\pm,||}=R_{J-1,\nu}(\xi_\pm)\left[\left(1\mp\frac{if_{4\pm}x^2}{\bar{c}}+\frac{x}{8\bar{c}}\right)\mathbf{v}_{J\sigma^+\nu}+A_{J-1,\nu}^{\nu-1}\mathbf{v}_{J\sigma^+,\nu-1}+A_{J-1,\nu}^{\nu+1}\mathbf{v}_{J\sigma^+,\nu+1}\right]-\\
&\frac{i}{\bar{c}}\frac{\xi_\pm}{\sqrt{1+\xi_\pm^2}}\mathbf{u}_{1J\sigma^+\nu}\Big|_{\xi=\xi_\pm}+\frac{1}{2\bar{c}}\frac{\xi_\pm^2}{1+\xi_\pm^2}\mathbf{u}_{2J\sigma^+\nu}\Big|_{\xi=\xi_\pm}+{\cal O}\left(\frac{1}{\bar{c}^2}\right)
\end{split}\end{equation}
and
\begin{equation}\begin{split}\label{sigma- on mirror}
&\mathbf{E}_{J\sigma^-\nu}^+\Big|_{\mathbf{S}_\pm,||}=R_{J+1,\nu}(\xi_\pm)\left[\left(1\mp\frac{if_{4\pm}x^2}{\bar{c}}-\frac{x}{8\bar{c}}\right)\mathbf{v}_{J\sigma^-\nu}+A_{J+1,\nu}^{\nu-1}\mathbf{v}_{J\sigma^-,\nu-1}+A_{J+1,\nu}^{\nu+1}\mathbf{v}_{J\sigma^-,\nu+1}\right]+\\
&\frac{i}{\bar{c}}\frac{\xi_\pm}{\sqrt{1+\xi_\pm^2}}\mathbf{u}_{1J\sigma^-\nu}\Big|_{\xi=\xi_\pm}+\frac{1}{2\bar{c}}\frac{\xi_\pm^2}{1+\xi_\pm^2}\mathbf{u}_{2J\sigma^-\nu}\Big|_{\xi=\xi_\pm}+{\cal O}\left(\frac{1}{\bar{c}^2}\right).
\end{split}\end{equation}
For $\mathbf{E}_{J\sigma\nu}^-$ we must again replace $\xi_\pm$ everywhere by $-\xi_\pm$ and $f_{4\pm}$ by $-f_{4\pm}$. The $A_{J\nu}^s$ are expansion coefficients for the angular spheroidal functions~\cite{Zeppenfeld09}. We have $A_{J+1,\nu}^{\nu\pm1}={\cal O}(\frac{1}{\bar{c}})$ so that to lowest order in $\frac{1}{\bar{c}}$, $\mathbf{E}_{J\sigma\nu}^\pm|_{\mathbf{S}_\pm,||}$ is simply proportional to $\mathbf{v}_{J\sigma\nu}$, motivating the choice of $\mathbf{v}_{J\sigma\nu}$.

Eqs.~(\ref{sigma+ on mirror}) and (\ref{sigma- on mirror}) have the form
\begin{equation}\label{E|S expansion}
\mathbf{E}_{J\sigma\nu}^{\pm_1}\Big|_{\mathbf{S}_{\pm_2},||}=\sum_{\sigma'\nu'}\mathbf{v}_{J\sigma'\nu'}a_{\nu'\nu}^{\pm_1,\pm_2,J,\sigma',\sigma}
\end{equation}
with appropriate coefficients $a_{\nu'\nu}^{\pm_1,\pm_2,J,\sigma',\sigma}$. The subscripts on the $\pm$ signs denote two independent choices of $+$ or $-$. Eqs.~(\ref{sigma+ on mirror}) and (\ref{sigma- on mirror}) contain factors of $x$ as part of the coefficients of the functions $\mathbf{v}_{J\sigma\nu}$. These can be removed using the relations
\begin{equation}\label{xv+}
x\,\mathbf{v}_{J\sigma^+\nu}=-(\nu+J-1)\mathbf{v}_{J\sigma^+,\nu-1}+(2\nu+J)\mathbf{v}_{J\sigma^+\nu}-(\nu+1)\mathbf{v}_{J\sigma^+,\nu+1}
\end{equation}
and
\begin{equation}\label{xv-}
x\,\mathbf{v}_{J\sigma^-\nu}=-(\nu+J+1)\mathbf{v}_{J\sigma^-,\nu-1}+(2\nu+J+2)\mathbf{v}_{J\sigma^-\nu}-(\nu+1)\mathbf{v}_{J\sigma^-,\nu+1},
\end{equation}
making the coefficients $a_{\nu'\nu}^{\pm_1,\pm_2,J,\sigma',\sigma}$ independent of $x$ and $\phi$.

The boundary condition for $\mathbf{E}$ can now be written as
\begin{equation}\begin{split}\label{boundary condition}
\mathbf{E}\Big|_{\mathbf{S}_\pm,||}&=\sum_{\sigma\nu}\left(b_{J\sigma\nu}^+\mathbf{E}_{J\sigma\nu}^+\Big|_{\mathbf{S}_\pm,||}+b_{J\sigma\nu}^-\mathbf{E}_{J\sigma\nu}^-\Big|_{\mathbf{S}_\pm,||}\right)\\
&=\sum_{\sigma'\nu'}\mathbf{v}_{J\sigma'\nu'}\left[\sum_{\sigma\nu}\left(a_{\nu'\nu}^{+,\pm,J,\sigma',\sigma}b_{J\sigma\nu}^++a_{\nu'\nu}^{-,\pm,J,\sigma',\sigma}b_{J\sigma\nu}^-\right)\right]=0.
\end{split}\end{equation}
Due to cylindrical symmetry, modes of different $J$ do not couple, so we have restricted our attention to a single $J$. The $\mathbf{v}_{J\sigma\nu}$, considered as vector functions of $x$ and $\phi$, are linearly independent. As a result, the expression in the square bracket in Eq.~(\ref{boundary condition}) must be zero for each value of $\sigma'$ and $\nu'$.

We define the following matrices and vectors,
\begin{equation}\label{A and b}
A_{\sigma'\sigma}^{\pm_1,\pm_2}=(a_{\nu'\nu}^{\pm_1,\pm_2,J,\sigma',\sigma})_{\nu'\nu},\hspace{1cm}
A^{\pm_1,\pm_2}=\begin{pmatrix} A_{\sigma^+\sigma^+}^{\pm_1,\pm_2} & A_{\sigma^+\sigma^-}^{\pm_1,\pm_2}\\ A_{\sigma^-\sigma^+}^{\pm_1,\pm_2} & A_{\sigma^-\sigma^-}^{\pm_1,\pm_2}\end{pmatrix},\hspace{6mm}\mbox{and}\hspace{6mm}
\mathbf{b}_{\sigma}^\pm=(b_{J\sigma\nu}^\pm)_\nu.
\end{equation}
The matrix $A^{\pm_1,\pm_2}$ maps the traveling wave described by the vector $\mathbf{b}_{\sigma}^{\pm_1}$ of coefficients of wavefunctions $\mathbf{E}_{J\sigma\nu}^{\pm_1}$ onto a set of coefficients of surface functions $\mathbf{v}_{J\sigma\nu}$ describing the electric field component parallel to the surface at $\mathbf{S}_{\pm_2}$. This finally allows us to write the boundary conditions as matrix equations,
\begin{equation}
A^{+,\pm}\begin{pmatrix} \mathbf{b}_{\sigma^+}^+\\ \mathbf{b}_{\sigma^-}^+\end{pmatrix}+A^{-,\pm}\begin{pmatrix} \mathbf{b}_{\sigma^+}^-\\ \mathbf{b}_{\sigma^-}^-\end{pmatrix}=0.
\end{equation}
Solving for $\mathbf{b}_\sigma^\mp$ in terms $\mathbf{b}_\sigma^\pm$ using the boundary condition imposed by the mirror at $\xi_\pm$, we find the round-trip matrix for the resonator to be given by
\begin{equation}\label{round-trip matrix}
A=(A^{+,-})^{\hspace{-1mm}^{-1}}A^{-,-}\,(A^{-,+})^{\hspace{-1mm}^{-1}}A^{+,+}.
\end{equation}

The eigenvectors of $A$ are vectors of coefficients $b_{J\sigma\nu}^+$ of resonator eigenmodes, the corresponding eigenvalues are the round-trip phase shifts. To lowest order in $\frac{1}{\bar{c}}$, the matrix $A$ is diagonal. The resonator eigenmodes are therefore of the form $b_{J\sigma^\pm\nu}^+\mathbf{E}_{J\sigma^\pm\nu}^++b_{J\sigma^\pm\nu}^-\mathbf{E}_{J\sigma^\pm\nu}^-$ and the corresponding round-trip phase shift is equal to
\begin{equation}\label{paraxial phase shift}
\lambda_{J\sigma^\pm\nu}=e^{2i\left(\bar{c}(\xi_+-\xi_-)-(2\nu+(J\mp1)+1)(\arctan(\xi_+)-\arctan(\xi_-))+{\cal O}\left(\frac{1}{\bar{c}}\right)\right)}.
\end{equation}
This result is equivalent to the one obtained by considering a resonator in the framework of the paraxial approximation.

\section{First-Order Corrections to the Round-Trip Phase Shift}\label{section correction}
First-order corrections to the round-trip phase shift of Eq.~(\ref{paraxial phase shift}) can be obtained via degenerate perturbation theory. We begin by determining those eigenmodes of the round-trip matrix $A$ which are degenerate to lowest order in $\frac{1}{\bar{c}}$. As can be seen from Eq.~(\ref{paraxial phase shift}), the lowest-order eigenvalues corresponding to the modes $\mathbf{E}_{J\sigma^+\nu}$ and $\mathbf{E}_{J\sigma^-,\nu-1}$ are the same for arbitrary $J$ and $\nu$. This reflects the fact that, within the paraxial approximation, the resonator eigenfrequency is independent of the polarization. Additionally, for a cavity geometry such that $\frac{\arctan(\xi_1)-\arctan(\xi_2)}{\pi/2}$ is a rational number, say $\frac{p}{n}$ with relatively prime integers $p$ and $n$, the lowest-order eigenvalues corresponding to the modes $\mathbf{E}_{J\sigma\nu}$ and $\mathbf{E}_{J\sigma,\nu+n}$ are the same for arbitrary $J$, $\nu$ and $\sigma$.

Lowest-order degeneracy between modes is only relevant if the modes are coupled by first-order matrix elements. As can be seen from Eqs.~(\ref{sigma+ on mirror}) and (\ref{sigma- on mirror}) together with Eqs.~(\ref{u1+}-\ref{u2-}), a mode $\mathbf{E}_{J\sigma^+\nu}$ in general couples to the modes $\mathbf{E}_{J\sigma^+,\nu\pm1}$ as well as to the modes $\mathbf{E}_{J\sigma^-\nu'}$ with $\nu'=\nu-2$, $\nu-1$ and $\nu$. A mode $\mathbf{E}_{J\sigma^-\nu}$ couples to the modes $\mathbf{E}_{J\sigma^-,\nu\pm1}$ as well as to the modes $\mathbf{E}_{J\sigma^+\nu'}$ with $\nu'=\nu$, $\nu+1$ and $\nu+2$. For nonzero $f_{4\pm}$ coefficient, the modes $\mathbf{E}_{J\sigma^+\nu}$ and $\mathbf{E}_{J\sigma^-\nu}$ additionally couple to the modes $\mathbf{E}_{J\sigma^+\nu\pm2}$ and $\mathbf{E}_{J\sigma^-\nu\pm2}$, respectively.

From this analysis, we see that the degeneracy between the modes $\mathbf{E}_{J\sigma^+\nu}$ and $\mathbf{E}_{J\sigma^-,\nu-1}$ must be taken into account. On the other hand, a degeneracy between the modes $\mathbf{E}_{J\sigma^+\nu}$ and $\mathbf{E}_{J\sigma^+,\nu+n}$ due to resonator geometry is only relevant for $n=1$ and for $n=2$, $f_{4\pm}\ne0$. For these two special cases, an infinite set of degenerate modes are coupled by first-order off-diagonal matrix elements, and finding the first-order corrections to the eigenvalues of $A$ is significantly more difficult. We therefore exclude these two cases from further analysis.

Disregarding the two cases $n=1$ and $n=2$, $f_{4\pm}\ne0$, the first-order corrections to the eigenvalues of $A$ can be obtained from the two-by-two submatrix of $A$ which corresponds to the subspace spanned by $\mathbf{E}_{J\sigma^+\nu}$ and $\mathbf{E}_{J\sigma^-,\nu-1}$. We denote this matrix as the two-by-two round-trip matrix. Since the four matrices $A^{\pm_1,\pm_2}$ are all diagonal to lowest order in $\frac{1}{\bar{c}}$, the two-by-two round-trip matrix is given to first order in $\frac{1}{\bar{c}}$ by a product as in Eq.~(\ref{round-trip matrix}), except that each of the matrices $A^{\pm_1,\pm_2}$ is replaced by an appropriate two-by-two submatrix. Specifically, these four submatrices are made up of the coefficients $a_{ij}^{\pm_1,\pm_2}$ with $i,j\in\{0,1\}$, given by
\begin{equation}\begin{split}
a_{00}^{\pm_1,\pm_2}=a_{\nu\nu}^{\pm_1,\pm_2,J,\sigma^+,\sigma^+},\hspace{2cm}a_{10}^{\pm_1,\pm_2}=a_{\nu-1,\nu}^{\pm_1,\pm_2,J,\sigma^-,\sigma^+},\\
a_{01}^{\pm_1,\pm_2}=a_{\nu,\nu-1}^{\pm_1,\pm_2,J,\sigma^+,\sigma^-},\hspace{2cm}a_{11}^{\pm_1,\pm_2}=a_{\nu-1.\nu-1}^{\pm_1,\pm_2,J,\sigma^-,\sigma^-}.
\end{split}\end{equation}
The coefficients $a_{ij}^{+,\pm}$, obtained by applying Eqs.~(\ref{xv+}) and (\ref{xv-}) to Eqs.~(\ref{sigma+ on mirror}) and (\ref{sigma- on mirror}) and comparing with Eq.~(\ref{E|S expansion}), are given explicitly by
\begin{displaymath}\begin{split}
a_{00}^{+,\pm}&=R_{J-1,\nu}(\xi_\pm)\left(1\mp\frac{if_{4\pm}(6\nu^2+6J\nu+J^2+J)}{\bar{c}}+\frac{2\nu+J}{8\bar{c}}\right)+\\
&-\frac{i}{\bar{c}}\frac{\xi_\pm}{\sqrt{1+\xi_\pm^2}}\left[\frac{\nu}{2}\,R_{J,\nu-1}(\xi_\pm)-\frac{\nu+J}{2}\,R_{J\nu}(\xi_\pm)\right]-\frac{2\nu+J}{4\bar{c}}\frac{\xi_\pm^2}{1+\xi_\pm^2}R_{J-1,\nu}(\xi_\pm)+{\cal O}\left(\frac{1}{\bar{c}^2}\right),
\\\\
a_{10}^{+,\pm}&=\frac{i}{2\bar{c}}\frac{\xi_\pm}{\sqrt{1+\xi_\pm^2}}\Big[R_{J,\nu-1}(\xi_\pm)-R_{J\nu}(\xi_\pm)\Big]+\frac{1}{2\bar{c}}\frac{\xi_\pm^2}{1+\xi_\pm^2}R_{J-1,\nu}(\xi_\pm)+{\cal O}\left(\frac{1}{\bar{c}^2}\right),
\end{split}\end{displaymath}
\begin{equation}\begin{split}
a_{11}^{+,\pm}&=R_{J+1,\nu-1}(\xi_\pm)\left(1\mp\frac{if_{4\pm}(6\nu^2+6J\nu+J^2-J)}{\bar{c}}-\frac{2\nu+J}{8\bar{c}}\right)+\\
&\frac{i}{\bar{c}}\frac{\xi_\pm}{\sqrt{1+\xi_\pm^2}}\left[-\frac{\nu+J}{2}\,R_{J,\nu-1}(\xi_\pm)+\frac{\nu}{2}\,R_{J\nu}(\xi_\pm)\right]-\frac{2\nu+J}{4\bar{c}}\frac{\xi_\pm^2}{1+\xi_\pm^2}R_{J+1,\nu-1}(\xi_\pm)+{\cal O}\left(\frac{1}{\bar{c}^2}\right),
\\
\mbox{and}&
\\
a_{01}^{+,\pm}&=\frac{i\nu(\nu+J)}{2\bar{c}}\frac{\xi_\pm}{\sqrt{1+\xi_\pm^2}}\Big[R_{J,\nu-1}(\xi_\pm)-R_{J\nu}(\xi_\pm)\Big]+\frac{\nu(\nu+J)}{2\bar{c}}\frac{\xi_\pm^2}{1+\xi_\pm^2}R_{J+1,\nu-1}(\xi_\pm)+{\cal O}\left(\frac{1}{\bar{c}^2}\right).
\end{split}\end{equation}
For $a_{ij}^{-,\pm}$ one can check that $a_{ij}^{-,\pm}=(a_{ij}^{+,\pm})^\star$ with $^\star$ denoting complex conjugation.

The half of the two-by-two round-trip matrix corresponding to the mirror at $\xi_\pm$ is given by
\begin{equation}\begin{split}\label{two by two mirror}
\begin{pmatrix} a_{00}^{\mp,\pm} & a_{01}^{\mp,\pm}\\ a_{10}^{\mp,\pm} & a_{11}^{\mp,\pm}\end{pmatrix}^{-1}\begin{pmatrix} a_{00}^{\pm,\pm} & a_{01}^{\pm,\pm}\\ a_{10}^{\pm,\pm} & a_{11}^{\pm,\pm}\end{pmatrix}=&\\
\frac{1}{a_{00}^{\mp,\pm}a_{11}^{\mp,\pm}-a_{01}^{\mp,\pm}a_{10}^{\mp,\pm}}&\begin{pmatrix}a_{11}^{\mp,\pm}a_{00}^{\pm,\pm}-a_{01}^{\mp,\pm}a_{10}^{\pm,\pm} & a_{11}^{\mp,\pm}a_{01}^{\pm,\pm}-a_{01}^{\mp,\pm}a_{11}^{\pm,\pm} \\ -a_{10}^{\mp,\pm}a_{00}^{\pm,\pm}+a_{00}^{\mp,\pm}a_{10}^{\pm,\pm} & -a_{10}^{\mp,\pm}a_{01}^{\pm,\pm}+a_{00}^{\mp,\pm}a_{11}^{\pm,\pm}\end{pmatrix}\\
\end{split}\end{equation}
Surprisingly, the off-diagonal elements in Eq.~(\ref{two by two mirror}) are zero to first order in $\frac{1}{\bar{c}}$. This is due to the fact that the complex phase of the three functions $i(R_{J,\nu-1}(\xi)-R_{J\nu}(\xi))$, $R_{J+1,\nu-1}(\xi)$ and $R_{J-1,\nu}(\xi)$ is equal to lowest order in $\frac{1}{\bar{c}}$, i.e.
\begin{equation}
\arg(i(R_{J,\nu-1}(\xi)-R_{J\nu}(\xi)))=\arg(R_{J+1,\nu-1}(\xi))+{\cal O}\left(\frac{1}{\bar{c}}\right)=\arg(R_{J-1,\nu}(\xi))+{\cal O}\left(\frac{1}{\bar{c}}\right).
\end{equation}
This in turn results in equal complex phase to lowest order in $\frac{1}{\bar{c}}$ for the four coefficients $a_{ij}^{\pm_1,\pm_2}$ with fixed $\pm_1$ and $\pm_2$. Since $a_{ij}^{-,\pm}$ is the complex conjugate of $a_{ij}^{+,\pm}$, the lowest-order term of $a_{ij}^{-,\pm}a_{i'j'}^{+,\pm}$ is real for arbitrary $i$, $j$, $i'$ and $j'$. Since $a_{ij}^{+,\pm}={\cal O}(1)$ for $i=j$ and $a_{ij}^{+,\pm}={\cal O}(\frac{1}{\bar{c}})$ for $i\ne j$, we have
\begin{equation}\begin{split}
a_{00}^{\mp,\pm}a_{10}^{\pm,\pm}-a_{10}^{\mp,\pm}a_{00}^{\pm,\pm}&=a_{00}^{\mp,\pm}a_{10}^{\pm,\pm}-(a_{10}^{\pm,\pm}a_{00}^{\mp,\pm})^\star\\
&=a_{00}^{\mp,\pm}a_{10}^{\pm,\pm}-a_{00}^{\mp,\pm}a_{10}^{\pm,\pm}\times\left(1+{\cal O}\left(\frac{1}{\bar{c}}\right)\right)={\cal O}\left(\frac{1}{\bar{c}^2}\right)
\end{split}\end{equation}
and similarly
\begin{equation}
a_{11}^{\mp,\pm}a_{01}^{\pm,\pm}-a_{01}^{\mp,\pm}a_{11}^{\pm,\pm}={\cal O}\left(\frac{1}{\bar{c}^2}\right).
\end{equation}
For the diagonal elements in Eq.~(\ref{two by two mirror}), we note that $a_{01}^{\pm_1,\pm_2}a_{10}^{\pm_1,\pm_2}={\cal O}(\frac{1}{\bar{c}^2})$, so the top left element, which we denote $\lambda_{J\sigma^+\nu}^\pm$, is given by $a_{0,0}^{\pm,\pm}/a_{0,0}^{\mp,\pm}$, and the bottom right element, which we denote $\lambda_{J\sigma^-,\nu-1}^\pm$, is given by $a_{1,1}^{\pm,\pm}/a_{1,1}^{\mp,\pm}$.

The fact that the off-diagonal elements in Eq.~(\ref{two by two mirror}) are zero through first order in $\frac{1}{\bar{c}}$ means that the mirrors do not couple the modes $\mathbf{E}_{J\sigma^+\nu}$ and $\mathbf{E}_{J\sigma^-,\nu-1}$ to first order. As a result, $\mathbf{E}_{J\sigma^+\nu}$ and $\mathbf{E}_{J\sigma^-,\nu-1}$ remain lowest-order resonator eigenmodes when taking into account first-order corrections, despite being degenerate at lowest order. The round-trip phase shifts for the two modes are given by the products $\lambda_{J\sigma^+\nu}=\lambda_{J\sigma^+\nu}^+\lambda_{J\sigma^+\nu}^-$ and $\lambda_{J\sigma^-,\nu-1}=\lambda_{J\sigma^-,\nu-1}^+\lambda_{J\sigma^-,\nu-1}^-$. With a bit of algebra we obtain the central result,
\begin{equation}\begin{split}\label{lambda_J,sigma+,nu}
&\lambda_{J\sigma^+\nu}^\pm=\exp\bigg[\pm2i\bigg(\bar{c}\,\xi_\pm-(2\nu+J)\arctan(\xi_\pm)-\\
&\frac{1}{\bar{c}}\left(\frac{\xi_\pm}{1+\xi_\pm^2}\nu(\nu+J)\pm f_{4\pm}\Big(6\nu(\nu+J)+J(J+1)\Big)\right)+{\cal O}\left(\frac{1}{\bar{c}^2}\right)\bigg)\bigg]
\end{split}\end{equation}
and
\begin{equation}\begin{split}\label{lambda_J,sigma-,nu-1}
&\lambda_{J\sigma^-,\nu-1}^\pm=\exp\bigg[\pm2i\bigg(\bar{c}\,\xi_\pm-(2\nu+J)\arctan(\xi_\pm)-\\
&\frac{1}{\bar{c}}\left(\frac{\xi_\pm}{1+\xi_\pm^2}\nu(\nu+J)\pm f_{4\pm}\Big(6\nu(\nu+J)+J(J-1)\Big)\right)+{\cal O}\left(\frac{1}{\bar{c}^2}\right)\bigg)\bigg].
\end{split}\end{equation}

Note that the first line of both Eqs.~(\ref{lambda_J,sigma+,nu}) and (\ref{lambda_J,sigma-,nu-1}) is simply the paraxial result of Eq.~(\ref{paraxial phase shift}). The term proportional to $f_{4\pm}$ has been calculated previously~\cite{Visser05}, and is simply equal to the phase shift corresponding to the intensity weighted average position shift of the mirror due to a change in $f_{4\pm}$.

\section{Discussion of the Results}
The resonator round-trip phase shift of Eqs.~(\ref{lambda_J,sigma+,nu}) and (\ref{lambda_J,sigma-,nu-1}) allows the calculation of the exact next-order correction to the eigenfrequencies of a Fabry-Perot resonator with perfectly conducting mirrors. Beyond providing a bound for the accuracy of the paraxial approximation, the correction term has a number of physical implications. One of the most interesting is the prediction of spin-orbit coupling in a resonator. For two modes with equal $\nu$ and equal orbital angular momentum $m=J\mp1$ but opposite spin $\sigma^\pm$, we find a difference in round-trip phase shift equal to
\begin{equation}\label{spin-orbit coupling}
\frac{\lambda_{m+1,\sigma^+\nu}^\pm}{\lambda_{m-1,\sigma^-\nu}^\pm}=\exp\left(\pm\frac{2i}{\bar{c}}\frac{\xi_\pm}{1+\xi_\pm^2}\times m\right)=\exp\left(\frac{2im}{kR_\pm}\right).
\end{equation}
The resonator eigenfrequency is therefore equal to the sum of a term independent of spin and a term proportional to the product of spin and orbital angular momentum, i.e. spin and orbital angular momentum are coupled.

The correction term for the resonator round-trip phase shift fundamentally alters the notion of degeneracy of modes in a resonator. In the absence of this term, all modes with equal $2\nu+m$ are degenerate, independent of polarization. This allows significant arbitrariness in the choice of eigenmode basis for a resonator, i.e. Hermite-Gaussian modes, Laguerre-Gaussian modes and many other linear combinations are equally valid. For a resonator geometry such that $\frac{\arctan(\xi_1)-\arctan(\xi_2)}{\pi/2}=\frac{p}{n}$, additional degeneracy results, making, e.g., a confocal resonator an apparently ideal choice in applications where a cavity with many degenerate modes is desired.

Taking the corrections into account, this picture breaks down completely. Degeneracy between the various modes is almost completely lifted. Circularly polarized Laguerre-Gaussian modes are clearly singled out as the fundamental resonator modes, consistent with the cylindrical symmetry. This removal of degeneracy is the ultimate situation for spherical mirror resonators. However, the presence of the $f_{4\pm}$ term in the round-trip phase shift opens completely new possibilities via the capability to manufacture aspherical mirrors. For a choice of $f_{4\pm}=\mp\frac{1}{6}\,\xi_\pm/(1+\xi_\pm^2)$, the correction term for the round-trip phase shift is significantly reduced, leaving only a term which is independent of $\nu$ and mainly quadratic in $J$. This allows for the construction of resonators with a degeneracy of a large number of modes reestablished. Note however that this possibility does not include confocal resonators. Such resonators fall in the class of resonators with $n=1$ as defined above, causing the derivation of Eqs.~(\ref{lambda_J,sigma+,nu}) and (\ref{lambda_J,sigma-,nu-1}) in section \ref{section correction} to fail. Together with the fact that they lie at the edge of the zone of stability, confocal resonators turn out to be among the worst possible choices for a degenerate resonator.


Last but not least, we consider the prospect of observing the effects of the correction term experimentally. While this should be relatively easy for a resonator with dimensions on the order of the wavelength, an alternative highly interesting system would be an ultra-high-finesse macroscopic resonator in the optical domain. Although the expected frequency shift for the lowest-order modes of such a resonator is near the resolution limit and is likely to be swamped by mirror imperfections, the quadratic dependence of the correction term on the mode indices $J$ and $\nu$ should allow the correction term to be relatively easily observed for higher-order modes. Verification of the present results in such a system would constitute a precision test of diffraction and propagation in a resonator.

\section*{Acknowledgments}
We thank Michael Motsch for interesting discussions. Support by the Deutsche Forschungsgemeinschaft via the excellence cluster "Munich Centre for Advanced Photonics" and via EuroQUAM (Cavity-Mediated Molecular Cooling) is acknowledged.


\begin{thebibliography}{10}

\bibitem{Schuster08}
I. Schuster , A. Kubanek, A. Fuhrmanek, T. Puppe, P.W.H. Pinkse, K. Murr, and G. Rempe,
"Nonlinear spectroscopy of photons bound to one atom,"
Nat. Phys. {\bf 4}, 382-385 (2008).

\bibitem{Kuhn02}
A. Kuhn, M. Hennrich, and G. Rempe,
"Deterministic single-photon source for distributed quantum networking,"
Phys. Rev. Lett. {\bf 89}, 067901 (2002).

\bibitem{Lazutkin68}
V.F. Lazutkin,
"An equation for the natural frequencies of a nonconfocal resonator with cylindrical mirrors which takes mirror aberration into account,"
Opt. Spectr. {\bf 24}, 236 (1968).

\bibitem{Laabs99}
H. Laabs and A.T. Friberg,
"Nonparaxial Eigenmodes of Stable Resonators,"
IEEE J Quanum Electron. {\bf 35}, 198-207 (1999).

\bibitem{Visser05}
J. Visser and G. Nienhuis,
"Spectrum of an optical resonator with spherical aberration,"
J. Opt. Soc. Am. A {\bf 22}, 2490-2497 (2005).

\bibitem{Zomer07}
F. Zomer, V. Soskov, and A. Variola,
"On the nonparaxial modes of two-dimensional nearly concentric resonators,"
Appl. Opt. {\bf 46}, 6859-6866 (2007).

\bibitem{Lax75}
M. Lax, W.H. Louisell and W.B. McKnight,
"From Maxwell to paraxial wave optics,"
Phys. Rev. A {\bf 11}, 1365-1370 (1975).

\bibitem{Zimmerer63}
R.W. Zimmerer,
"Spherical Mirror Fabry-Perot Resonators,"
IEEE Trans. Microwave Theory Tech. {\bf 11}, 371-379 (1963).

\bibitem{Specht65}
W.A. Specht Jr.,
"Modes in Spherical-Mirror Resonators,"
J. Appl. Phys. {\bf 36}, 1306-1313 (1965).

\bibitem{Flammer57}
C. Flammer. {\it Spheroidal Wave Functions}. (Stanford University Press, Stanford, Calif., 1957).

\bibitem{Zeppenfeld09}
M. Zeppenfeld,
"Solutions to Maxwell's equations using spheroidal coordinates,"
New J. Phys. {\bf 11}, 073007 (2009).

\bibitem{Abramowitz+Stegun}
M. Abramowitz, I.A. Stegun. {\it Handbook of Mathematical Functions}. (Dover, New York, 1964).

\end{thebibliography}
\end{document}